# Control of Förster energy transfer in vicinity of metallic surfaces and hyperbolic metamaterials

T. U. Tumkur[a], J. K. Kitur[a], C. E. Bonner[a], A. N. Poddubny[b,c] E. E. Narimanov[d], and M. A. Noginov[a]*

[a] *Center for Materials Research, Norfolk State University, Norfolk, VA 23504*

*mnoginov@nsu.edu

[b] *ITMO University, St. Petersburg 197101, Russia*

[c] *Ioffe Physical-Technical Institute, St. Petersburg 194021, Russia*

[d] *Birck Nanotechnology Center, Department of Electrical and Computer Engineering, Purdue University, West Lafayette, IN 47907*

**Abstract:**

Optical cavities, plasmonic structures, photonic band crystals, interfaces, as well as, generally speaking, any photonic media with homogeneous or spatially inhomogeneous dielectric permittivity (including metamaterials) have local densities of photonic states, which are different from that in vacuum. These modified density of states environments are known to control both the rate and angular distribution of spontaneous emission. In the present study, we ask the question whether the proximity to metallic and metamaterial surfaces can affect other physical phenomena of fundamental and practical importance. We show that the same substrates and the same nonlocal dielectric environments that boost spontaneous emission, also inhibit Förster energy transfer between donor and acceptor molecules doped into a thin polymeric film. This finding correlates with the fact that in dielectric media, the rate of spontaneous emission is proportional to the index of refraction *n,* while the rate of the donor-acceptor energy transfer (in solid solutions with random distribution of acceptors) is proportional to $n^{-1.5}$. This heuristic correspondence suggests that other classical and quantum phenomena, which in regular dielectric media depend on *n*, can also be controlled with custom-tailored metamaterials, plasmonic structures, and cavities.

## 1. Introduction

Metamaterials – engineered composite materials containing subwavelength inclusions with tailored shapes, sizes, mutual arrangements and orientations[1,2]– fascinate scientists and engineers by their unparalleled responses to electromagnetic waves[3-7]. Thus, metamaterials with hyperbolic dispersion, whose dielectric permittivities in orthogonal directions have different signs[6,8,9,10] propagate waves with nearly unlimited wavevectors and have a broad-band singularity of the local density of photonic states[11]. The latter phenomenon, in spirit of the Fermi's golden rule, can control spontaneous emission[12-15] and reflectance of roughened metamaterials' surfaces[16,17].

In this study, we have researched the effect of the dielectric environment, including the local density of photonic states $\rho$, on Förster energy transfer between donor and acceptor molecules[18]. The literature has several contradictory theoretical and experimental studies of the effect of the local density of photonic states on the Förster energy transfer. Thus, its rate was claimed to be dependent on[19-23] or independent[24-27] of the photonic environment. Experimentally, a high density of photonic states was reported to enhance the energy transfer rate in cavities[20], modulate it[21,22], or leave it unaffected[24-27] in vicinity of mirrors and planar interfaces, or inhibit it in plasmonic structures[28]. This broad range of claims and opinions makes our findings outlined below particularly important.

Förster energy transfer typically occurs between donors and acceptors situated in close proximity (<<λ). This interaction of Coulomb nature is usually mediated by a polarizable medium of the host matrix. At dipole-dipole character of the energy transfer, fixed positions of donors and acceptors, and random spatial distributions of acceptors around donors, the emission kinetics of donors *I(t)* (excited by short laser pulses) is given by[18]

$$I(t) = I_0 \exp\left(-(A+W)t - \gamma\sqrt{t}\right), \quad (1)$$

where $I_0$ is the initial emission intensity, $A$ and $W$ are the radiative and non-radiative emission decay rates of donors, $t$ is the time, $\gamma \approx 6.28 R_0^3 N / \sqrt{\tau_0}$ is the energy transfer constant, $N$ is the concentration of acceptors, and $\tau_0 = (A+W)^{-1}$ is the decay-time of donors in the absence of acceptors. Here

$$R_0 = \sqrt[6]{\frac{3}{2(2\pi)^5} \frac{\eta_0}{n^4} \int F(\bar{\nu})\sigma(\bar{\nu})\frac{d\bar{\nu}}{\bar{\nu}^4}} \quad (2)$$

is the characteristic distance at which the energy transfer rate in a pair of donor and acceptor is equal to the emission decay rate of the donor molecule, $\bar{\nu} = \omega/2\pi c$ is the frequency in cm$^{-1}$, $\omega$ is the angular frequency, $c$ is the speed of light, $F(\bar{\nu})$ is the normalized radiation spectrum of donors, $\sigma(\bar{\nu})$ is the absorption cross section spectrum of acceptors, $\eta_0 = A/(A+W)$ is the quantum yield of spontaneous emission of donors (in the absence of acceptors), and $n$ is the index of refraction.

## 2. Experimental Results

In this work, thin polymeric films doped by donor and acceptor molecules were deposited on a variety of substrates including glass (control sample), 200 nm Ag film on glass, 200 nm Au film on glass, as well as hyperbolic metamaterials. Known metamaterials with hyperbolic dispersion include ordered arrays of metallic nanowires[29,30] and multi-layered metal/dielectric or semiconductor lamellar thin film structures[13-15,31]. The latter morphology was employed in our studies. (See Section 5.1 for fabrication of hyperbolic metamaterials and study of their dispersion properties).

In an optimal pair of matching dyes, the emission band of a *donor* should spectrally overlap with the absorption band of an *acceptor*. This condition is satisfied in a Poly(methyl methacrylate) (PMMA) film co-doped with DCM dye molecules (donors), and DOTC dye molecules (acceptors), Fig. 1a. (See Section 5.2.) Trace 1 of Fig. 1b depicts the emission spectrum of such film deposited on glass and pumped at λ=400 nm into the absorption band of DCM. The emission band of the DOTC dye, clearly seen in this spectrum, manifests an efficient donor→acceptor energy transfer. Emission spectra of Fig. 1b, although useful for qualitative demonstrations, are not suitable for quantitative analysis of the donor→acceptor energy transfer on top of metallic and metamaterials substrates. This is because emission and excitation spectra in a mixture of donor and acceptor molecules depend not only on the energy transfer rate $\gamma$, but also on the absorption strength of donors, the radiative $A$ and non-radiative $W$ decay rates, as well as the directionality of emission of both donors and acceptors. As all these parameters can be sensitive to local densities of photonic states, evaluating the efficiency of the donor→acceptor energy transfer from the analysis of emission spectra becomes an extremely difficult task.

As it has been proven by numerous studies[32], a much better way to quantitatively characterize the efficiency of the Förster energy transfer is by recording and analyzing the emission kinetics of donors in presence of acceptors. Figure 2 depicts emission kinetics of DCM molecules embedded into PMMA film, excited at 392 nm with the second harmonic of 150 fs Ti:sapphire laser, as well as emission kinetics of DCM molecules (donors) and DOTC molecules (acceptors) co-doped into PMMA. (All three kinetics have been measured in dye doped PMMA films deposited on glass) The shortening of the emission kinetics of donors and delayed rise of the emission intensity of acceptors (Fig. 2), are the signature features of an efficient donor→acceptor energy transfer[18].

Dividing the emission decay kinetics of donors measured in a co-doped PMMA film (described by Eq. (1)) by that in the film doped with donor molecules only ($\propto \exp(-(A+W)t)$), we expected to single out the contribution of the Förster donor→acceptor energy transfer, $\propto \exp(-\gamma\sqrt{t})$. When we applied this procedure to emission kinetics 1 and 2 in Fig. 2, the resulting curve, plotted in the inset of Fig. 2 as {ln(-ln($I(t)$)) $vs$ ln($t$)}, had slope 0.6, which is close to 1/2 (expected of the function $\propto \exp(-\gamma\sqrt{t})$). This validates our method and gives us a powerful tool allowing to separate the Förster contribution to the emission kinetics (which nominally has slope 1/2) from all other radiative and non-radiative decay processes (which have slope 1). The corresponding energy transfer constant $\gamma$ can be determined by fitting the resultant experimental curve in inset of Fig. 2 by $\propto \exp(-\gamma\sqrt{t})$.

We have found that in polymeric films co-doped by donors and acceptors on top of Ag and Au films as well as lamellar metamaterials with Ag as the outermost layer, the rate of the Förster energy transfer is strongly inhibited. This is evidenced by the emission spectra of donor-acceptor mixtures pumped into the absorption band of donors, which lack emission of acceptors at ~710 nm (Fig. 1b, trace 2), as well as emission decay kinetics of donors, which are almost the same in the absence (trace 4) and in the presence (trace 5) of acceptors, Fig. 2.

By comparing traces 1 and 4 of Fig. 2, one can also see that the emission kinetics of (single-doped) donors on top of a metamaterial is shorter than that on top of glass. Following Refs. [11,13,15,25], we explain this phenomenon with a combination of spontaneous emission enhancement caused by the high local density of photonic states[11,33] (arguably a predominant effect) and nonradiative quenching of dye emission by metallic surface[34]. As the lifetime shortening is not uniform through the films' thickness, the emission decay kinetics showed deviation from a pure exponential form. The respective emission decay rates have been calculated as $I_0 / \int_0^\infty I(t)dt$.

At the same time, the emission kinetics of donors in the presence of acceptors is shorter on top of glass than on top of a metamaterial, compare traces 2 and 5 in Fig. 2. This, combined with nearly identical behavior of traces 4 and 5, is an unambiguous manifestation of the inhibition of the Förster energy transfer in the vicinity of lamellar metal/dielectric substrate. (Note, that the inhibition was weaker on top of a lamellar metamaterial, whose outermost layer was $MgF_2$, than on top a similar metamaterial, whose outmost layer was Ag.)

The values $\gamma$ obtained from polymeric films co-doped by donors and acceptors and corresponding emission decay rates $\tau_0^{-1}$ in films doped by donor molecules only have been measured in several tens of thin film samples (whose thickness ranged from 32 nm to 100 nm) deposited onto five types of substrates. The results of the measurements are summarized in Fig. 3a.

As one can see in Fig. 3a, the same metallic and hyperbolic metamaterial substrates and the same environments that enhance spontaneous emission decay (possibly with minor contribution from non-radiative decay), inhibit Förster energy transfer. This is the central experimental result of the present study. This observation correlates with the qualitative argument that, in regular dielectric media, the rate of spontaneous emission *A* (which we assume to be nearly $\propto \tau_0^{-1}$) is proportional to *n*, while $\gamma \propto n^{-1.5}$. The pairs of data points ($\gamma, \tau_0^{-1}$) measured on top of multiple metamaterial, metallic, and glass substrates are plotted in logarithmic coordinates Fig. 3b. The slope of the curve is equal to -1.8±0.42, in a fair agreement with the heuristic argument ($A \propto n$, $\gamma \propto n^{-1.5} \Rightarrow A \propto \gamma^{-1.5}$).

## 3. Theoretical Analysis and Discussion

### 3.1 Comparison with the literature

Although our findings are consistent with those of Ref. [28], in which the inhibition of the Förster energy transfer was observed in vicinity of plasmonic nanoparticles, it seemingly disagrees with several other reports, including Ref. [25], which claims an independence of the Förster energy transfer rate of the molecule-to-metal distance and the corresponding local density of photonic states. In the latter study, the distance between donor-acceptor pairs and the metallic mirror ranged between 60 nm and 270 nm, and the developed theoretical model predicted almost no dependence of the Förster energy transfer rate on the local density of states if the dielectric environment did not change on a scale smaller than the wavelength (~500 nm). In our work, the distances between donor-acceptor pairs and the metamaterial or metallic substrates were substantially smaller, ranging between 0 nm and 32 nm in thinner dye-doped films and between 0 nm and 95 nm in thicker films. This suggested that the effect of the substrate on the Förster energy transfer rate in our experiment could be larger than that in Ref. [25]. However, since the average molecule-to-substrate distances in our experiments were still large in comparison to the average donor-acceptor distance, *d*=4.4 nm, the effect the local density of photonic states on the rate of the Förster energy transfer had to be quantitatively evaluated as described below.

## 3.2 Theoretical formulation of the problem

The quasi-static result for the rate of the Förster energy transfer between donor and acceptor in an homogeneous medium reads

$$W_{DA} = W_0 \left(\frac{R_0}{r_{DA}}\right)^6 \quad (3)$$

where $r_{DA}$ is the distance between donor and acceptor molecules, $W_0 = A + W = \tau_0^{-1}$ is the decay rate of donors in the absence of acceptors, and the Förster radius $R_0$ (at which the rate of energy transfer is equal to the spontaneous emission rate) depends on intrinsic properties of the donor and acceptor as well as dielectric permittivity of the medium (Eq. 2). In an arbitrary medium characterized by the position-dependent dielectric permittivity $\varepsilon(r)$, Eq. (3) can be generalized to[35,36]

$$W_{DA}(r_A, r_B) = V Tr\left[\hat{G}(r_A, r_D)\hat{G}^+(r_A, r_D)\right] \quad (4)$$

where $\hat{G}(r,r')$ is the electromagnetic Green tensor satisfying the equation

$$\left[\Delta + \left(\frac{\omega}{c}\right)^2 \varepsilon(r)\right]\hat{G}(r,r') = -4\pi\left(\frac{\omega}{c}\right)^2 \left(\delta_{\alpha\beta} + \frac{1}{(\omega/c)^2 \varepsilon} \frac{\partial^2}{\partial x_\alpha \partial x_\beta}\right)\delta(r - r') \quad (5)$$

and evaluated at the donor emission frequency $\omega$; and $\alpha, \beta, \gamma$ are the Cartesian indices. The trace in Eq. 4 reflects averaging over the orientations of the donor and acceptor dipole matrix elements. The quantity $V$ in Eq. (4) depends only on the donor and acceptor properties, and all information about the electromagnetic environment is included into the Green function. The correspondence between Eq. (3) and Eq. (4) can be recovered in the case of a homogeneous medium with the dielectric permittivity $\varepsilon_b$, for the donor and acceptor molecules separated by the distance much smaller than the wavelength, $r_{DA} \ll c/\omega$. Under these assumptions, the Green function $G$ depends only on the relative coordinate $r_{DA} = r_D - r_A$ and is reduced to the

electrostatic dipole-dipole interaction term $G_0$

$$G_{0,\alpha\beta}(\mathbf{r}_{DA}) = \frac{3r_{DA,\alpha}r_{DA,\beta}}{\varepsilon_b r_{DA}^5} \frac{r_{DA}^2 \delta_{\alpha,\beta}}{} . \qquad (6)$$

Furthermore, the general result Eq. (4) reduces to Eq. (3), provided that $V = \varepsilon_b^2 W_0 R_0^6/6$. Equations 3 and 4 describe the rates of energy transfer between individual donor and acceptor molecules. For ensembles of donors and acceptors, there exists a distribution of the transfer rates depending on the intermolecular distance in the donor-acceptor pair. The decay kinetics of the given donor molecule is then determined by the transfer probabilities to different acceptors and strongly depends on the spatial distribution of the molecules. In the case when the acceptors are distributed homogeneously and independently in an infinitely large medium, the expression for the donor decay reads[18],

$$\begin{aligned}I(t) &= I_0 e^{-H(t)}, \\ H(t) &= \int_{r_{DA} > r_{min}} d^3 r_A n_A(r_A)\left(1 - e^{-W_{DA}(r_A, r_B)t}\right),\end{aligned} \qquad (7)$$

where the integration is performed over the spatial distribution of acceptors characterized with the concentration $n_A$. Here the radiative decay of the donors is neglected, cf. Eq. (7) and Eq. (1). The small sphere with the radius $r_{min}$ is excluded from the integration to account for the minimal distance between the molecules due to their finite size. For homogeneous medium, substituting Eq. (3) into Eq. (7) and neglecting $r_{min}$, we recover the energy transfer law Eq. (1) with $H(t) \propto \sqrt{t}$ [18].

*3.3 Effect of the silver substrate on the Förster energy transfer*

The calculated effect of the silver substrate on the energy transfer kinetics is presented in Fig. 4. The donor and acceptor molecules are embedded in the semi-infinite dielectric layer lying on top of the silver substrate. The dielectric-silver interface corresponds to the plane $z = 0$.

Acceptors are homogeneously distributed in a dielectric (polymer) film of thickness $d_{dye}$=30 nm deposited on top of silver ($0 < z_A < d_{dye}$); the concentration of acceptors is equal to $n_A$. In order to simplify the treatment, in our model the dielectric film with dye is covered by the same polymer. In the experimental samples the dye film is bounded by air from the top. Since the permittivity constrast between polymer and air is smaller than that between polymer and silver, the effect of the polymer/air boundary is weaker than the effect of the polymer/silver boundary, and this simplification should not significantly influence our results. Different curves in Fig. 4 correspond to two values of the height of the donor molecule $h \equiv z_D$ above the interface, $h$=2 nm and $h$=15 nm. The Green function has been evaluated according to Ref. [38]. At short times the decay law is exponential, with $H(t) \propto \sqrt{t}$, which is due to the finite value of $r_{min}$. This part of the decay curves is independent of the distance from the substrate provided it is larger than $r_{min}$. At larger times, the decay kinetics of the donor in the bulk is described by the $H(t) \propto \sqrt{t}$ law (cf. green and thin blue curves). For the donor located close to the substrate, the decay rate is slightly suppressed (blue curve). In principle, this suppression can be caused by a combination of the two reasons: (i) the donor at the surface has smaller number of acceptors in its vicinity than the donor in the bulk and (ii) the electromagnetic Green function near the surface is different from its bulk value, Eq. (6).

    The red dotted curve shows the contribution of the former effect only. It has been calculated using the bulk Green function that does not account for the presence of the substrate. Since this curve is very close to the blue solid curve, we conclude that the reduced number of available acceptors for the donors located near the surface of the polymeric film is the most important effect, while the Green function modification does not affect the energy transfer kinetics noticeably.

The analysis provided in Section 5.4 shows that, while the effect of modification of the Green function on the energy transfer rate is weak, the spatial modification pattern is quite complex. Thus, depending on the position of the acceptor, the Förster energy transfer can be either increased or suppressed.

We further demonstrate (in Section 5.4) that in the framework of our model, assuming interaction of individual donors with individual acceptors, the effects of the dielectric environment on the (i) density of photonic states and (ii) Förster energy transfer practically do not correlate with each other, with the latter effect being much weaker. Hence, from only the value of the density of states, it is not possible to deduce whether the energy transfer rate for a given donor-acceptor distance $r_{DA}$ will be enhanced or suppressed. This is explained by the fact that the energy transfer rate is determined by the electromagnetic modes with the wave vectors of the order of $1/r_{DA}$, while the density of states has contributions from all modes, in agreement with Ref. [25].

### 3.4 Collective dye-plasmon interactions

In the consideration above, we have assumed that each donor or acceptor molecule interacts with the electromagnetic modes of the system independently. However, for dense enough molecular ensemble and small inhomogeneous broadening, the collective effects may play an important role and result in formation of hybrid polaritonic excitations. Moreover, when the interaction between the excitonic modes of the donor (acceptor) ensembles becomes stronger than their dampings, one can attain the strong coupling regime, characterized by the anticrossings in the dispersion curves based on the emission, absorption and reflection spectra and the formation of hybridized exciton and plasmon eigenmodes[39,40].

Although not studied by us systematically, modification of absorption and/or excitation spectra of dye molecules in vicinity of metallic films and multilayered metal/dielectric structures is routinely observed in a variety of systems. As an example, the excitation spectra of the DOTC molecules in the PMMA film (co-doped by the DCM dye) deposited on glass and Ag/MgF$_2$ metamaterial are depicted in Fig. 5. The strong difference between the shapes of the two spectra suggests that the overlap integral (Eq. 2) and the rate of the Förster donor→acceptor energy transfer can be strongly modified by the nonlocal dielectric environments. Indeed, the energies, the oscillator strengths, and the lifetimes of the polaritonic modes are modified by the environment and are substantially different from those of the individual molecules. The detailed experimental investigation of this phenomenon as well as comprehensive theoretical study of the Förster energy transfer in nonlocal dielectric environments is the subject of an ongoing research effort to be published elsewhere.

*3.5 Potential applications*

Note that while some phenomena, including photosynthesis and photovoltaics, benefit from donor→acceptor energy transfer, others suffer from it badly. The latter include migration of electronic excitation to defects, which quenches the luminescence of luminophosphors, and self-quenching of luminescence in Nd$^{3+}$ doped laser crystals that limits maximal concentration of active ions and gain. Correspondingly, the possibility to inhibit the Förster energy transfer on demand can be very useful for a variety of applications.

**4. Summary**

To summarize, we show that the rate of the Förster energy transfer $\gamma$ in an ensemble of donor and acceptor molecules is inhibited on top of metallic and hyperbolic metamaterial substrates – the environments with high local densities of photonic states, which enhance spontaneous emission rates *A*. Although our theoretical model, describing interactions between *individual* independent donors and acceptors, predicts a weak reduction of the energy transfer on top of the substrate, the calculated effect is orders of magnitude smaller than the experimental one. We infer that the observed effect can be due to strong collectve interactions of the dye molecules and surface plasmon polaritons, which can affect the absorption and emission spectra and, correspondingly, change the rate of the Förster energy transfer.

## 5. Methods and Extended Data

*5.1 Fabrication and characterization of metamaterial and metallic film substrates*

The hyperbolic metamaterials samples, fabricated using physical vapor deposition technique, consisted of seven 25 nm layers of Ag and six 35 nm layers of $MgF_2$ (with Ag on the top) or nine layers of Ag and nine layers of $MgF_2$ (with $MgF_2$ on the top). The thickness of the layers was monitored with a quartz crystal sensor and later double-checked with the DekTak 6M profilometer.

The dielectric permittivities in the directions parallel ($\varepsilon_\parallel$) and perpendicular ($\varepsilon_\perp$) to the sample surface were calculated in the effective medium approximation[41] for a lamellar metal/dielectric structure,

$$\varepsilon_\parallel = f\varepsilon_m + (1-f)\varepsilon_d$$
$$\frac{1}{\varepsilon_\perp} = \frac{f}{\varepsilon_m} + \frac{(1-f)}{\varepsilon_d}, \qquad (8)$$

where $\varepsilon_m$ and *f* are the permittivity and the filling factor of a metal (0<*f*<1), and $\varepsilon_d$ is the permittivity of a dielectric. According to the spectra of the real parts of $\varepsilon_\parallel$ and $\varepsilon_\perp$ (calculated at *f*=0.39, corresponding to 25 nm of Ag and 35 nm of $MgF_2$), the material has a hyperbolic dispersion at $\lambda \geq 384$ nm, left inset of Fig. 6.

Experimentally, $\varepsilon_\parallel$ and $\varepsilon_\perp$ have been determined by measuring the angular reflectance profiles $R(\theta)$ in s and p polarizations (Fig. 6) and fitting them with the known formulas[30]

$$R = |r|^2 = \begin{cases} \left|\dfrac{\sin(\theta-\theta_t)}{\sin(\theta+\theta_t)}\right|^2, & \theta_t = \arcsin\left(\dfrac{\sin\theta}{\sqrt{\varepsilon_\parallel}}\right), & s\text{ polarization.} \\ \left|\dfrac{\varepsilon_\parallel \tan\theta_t - \tan\theta}{\varepsilon_\parallel \tan\theta_t + \tan\theta}\right|^2, & \theta_t = \arctan\sqrt{\dfrac{\varepsilon_\perp \sin^2\theta}{\varepsilon_\parallel \varepsilon_\perp - \varepsilon_\parallel \sin^2\theta}}, & p\text{ polarization.} \end{cases} \quad (9)$$

The values $\varepsilon_\parallel = -2.57 + i0.56$ and $\varepsilon_\perp = 5.56 + i0.64$ determined at $\lambda=543$ nm (the wavelength corresponding to the Förster energy transfer range in our experiment, see the main text), are reasonably close to the theoretical prediction, left inset of Fig. 6.

Furthermore, the calculated (normal incidence) reflectance spectrum is in a good agreement with the experimental reflectance spectrum measured at small incidence angle of 8 degrees, right inset of Fig. 6. We, thus, conclude that (i) our samples have hyperbolic dispersion in the whole visible and near infrared range that is of interest to our studies, and (ii) the dispersion can be reasonably well described in the effective medium approximation.

*5.2 Fabrication of dye-doped PMMA films*

The DCM ([2-[2-[4-(dimethylamino)phenyl]ethenyl]-6-methyl-4H- pyran-4-ylidene]-propanedinitrile) and DOTC (3-ethyl-2-[7-(3-ethyl-2(3H)-benzoxazolylidene)-1,3,5-heptatrienyl]-benzoxazolium iodide) dyes, along with Poly(methyl methacrylate) (PMMA, ~ 120000 avg. molecular weight) were dissolved in dichloromethane (in ultrasonic bath at 25°C) and spin-coated onto a variety of substrates discussed in Section 2. In our experiments, the concentration of DCM molecules was equal to $2\times10^{19}$ cm$^{-3}$ (33 mM or 10 g/l) and the concentration of DOTC molecules was equal to $1.2\times10^{19}$ cm$^{-3}$ (20 mM or 10 g/l). The film thicknesses, measured with Dektak 6M profilometer, ranged from 32 nm to 100 nm.

*5.3 Emission kinetics measurements and analysis*

The emission kinetics were detected with the Hamamatsu C5680 streak camera. The time resolution, determined by wide open slit of the streak camera and jitter of the laser, was ~80 ps. Appropriate combinations of color filters were used to selectively transmit emissions of DCM and DOTC and block scattered pumping light.

In the analysis of the emission kinetics, we relied on formula (1) derived under the assumption that the physical size of the medium is much larger than the average donor-acceptor distance $d$[18]. We infer that slight deviation of the experimentally measured slope in inset of Fig. 2 from 1/2, could be due to a relatively small thickness of the dye-doped films, which exceeded $d$=4.4 nm by approximately one order of magnitude.

*5.4 Theoretical studies*

Green function modification

Although, as shown in Section 3.3, the Green function modification practically does not affect the decay kinetics, it is still instructive to examine this effect in more detail. To this end, we analyze in Fig. 7 the dependence of the Förster energy transfer rate on the acceptor coordinate. The geometry of the problem is sketched in Fig. 7a.

Figure 7b presents the spatial map of the transfer rate of the donor positioned above the Ag mirror to different acceptors. Fig. 7d shows the same map for donor in the bulk. Both panels look quite similar. To reveal the difference, we examine in Fig. 7c the ratio between the rates, shown in two panels b and d. In particular, Fig. 7c presents the spatial map of the energy transfer rates normalized to the corresponding values evaluated in the same points in a bulk dielectric; it is obtained by dividing the data in Fig. 7b to those in Fig. 7d and substracting unity.

This figure demonstrates that while the relative modification of the energy transfer rate is weak, the modification pattern is quite complex. Depending on the position of the acceptor, the Förster energy transfer can be either increased or suppressed. Importantly, the energy transfer in the lateral direction, along the mirror surface, is suppressed for 10 nm < $r_{DA}$ < 60 nm, see blue areas in Fig. 7c.

This suppression can be well described within the image charge approximation for the Green function[42]. We consider as an example the transfer between copolarized donor and acceptor positioned on the same height from the surface. The relevant component of the Green tensor within the image charge approximation reads $G_{xx}(x_{DA},0,0) = 4/[|x_{DA}|^3(\varepsilon_{Ag} + \varepsilon_b)]$, where $\varepsilon_{Ag}$ is the dielectric permittivity of silver. This expression predicts strong enhancement of the transfer at the surface plasmon resonance, corresponding to the condition $\varepsilon_{Ag} + \varepsilon_b = 0$. However, the frequency range of the spectral overlap of emission of donors and absorption of acceptors is below the resonance, so $|\varepsilon_{Ag}| > |\varepsilon_b|$. As a result, the value of $G_{xx}$ is smaller than the bulk value $G_{0,xx} = 2/(|x_{DA}|^3 \varepsilon_b)$ and the transfer rate along the surface is suppressed, in agreement with Fig. 7c.

At the very large lateral distances, the energy transfer rate (normalized as in Fig. 7d) is increased as compared to the bulk value (red areas on both right hand and left hand sides of Fig. 7d). This is the surface plasmon-assisted transfer along the surface[43] – the effect, which is not captured by the quasi-static image charge model. The plasmon-assisted energy transfer becomes relatively large for lateral distances on the order of the inverse surface plasmon polariton (SPP) wave vector

$$k_{pl} = \frac{\omega}{c}\sqrt{\frac{\varepsilon_b \varepsilon_{Ag}}{\varepsilon_b + \varepsilon_{Ag}}}, \qquad (10)$$

for our parameters $\mathrm{Re}\{k_{pl}^{-1}\} \sim 60$ nm, in agreement with Fig. 7d.

However, even though the relative value of energy transfer rate with respect to the bulk value is increased, this practically does not affect the donor decay kinetics. The reason is that the absolute value of the Förster energy transfer rate for such large distance is quite small and the relative enhancement plays no role. In other words, the energy transfer rate is mostly determined by the short-range electromagnetic coupling between donor and acceptor, and the effect of the long-range plasmon assisted energy transfer is negligible. This effect can be more important at higher frequencies, which are closer to the surface plasmon resonance in silver. In fact, at higher frequencies, the surface plasmon wave vector $k_{pl}$ is larger, so the plasmons can contribute to the energy transfer stronger.

Figure 8 presents the dependence of the energy transfer rate on the donor-acceptor distance. The rate has been determined by averaging the data in Fig. 7 over the angular position of the acceptor. The calculation demonstrates that the rate deviates from the bulk decay law $1/r_{DA}^6$ only at very large distances, where the transfer is very weak.

Förster energy transfer and photonic local density of states

In this section, we analyze the effect of silver substrate on the Förster energy transfer rate and on the photonic local density of states, which determines the rate of spontaneous emission (Purcell effect)[44-46]. Figure 9 compares the effect of the donor position above the substrate $h$ on the Förster energy transfer with its effect on the Purcell enhancement factor $f_P$. Clearly, these two quantities are not correlated and the Purcell factor depends on $h$ much stronger than the energy transfer rate does. In fact, at $h$ equal to several nanometers, $f_P$ reaches the value of the order of hundred, constituting strong quenching of the molecule emission due to the Joule heating of the metal[18]. The Purcell factor, which can be approximated as[47]

$$f_p(h) = \frac{3}{8(\omega h/c)^3} \frac{\text{Im}\{\varepsilon_{Ag}\}}{|\varepsilon_b + \varepsilon_{Ag}|^2}, \qquad (11)$$

is very sensitive to the distance from the silver surface $h$. At the same time, the rate of the donor-acceptor energy transfer (calculated for fixed distances between donor and acceptor $r_{DA}=4$ nm) depends on the height $h$ weakly and nonmonotonously (see black dotted curve in Fig. 9). Therefore, the local dielectric environment affects the density of states and the Förster energy transfer rate in very different ways, with its effect on the energy transfer being much weaker. Hence, it is not possible to deduce whether the energy transfer rate for a given donor-acceptor distance $r_{DA}$ will be enhanced or suppressed based on the density of states value only. This is explained by the fact that the energy transfer rate is determined by the (short-range) electromagnetic modes with the wave vectors of the order of $1/r_{DA}$, while the density of states has contributions from all modes, in agreement with the findings of Ref. [25].

**Acknowledgments:** TUT, JKK, CEB, and MAN were supported by the NSF PREM grant DMR-1205457, NSF IGERT grant DGE-0966188, and AFOSR grant FA9550-09-1-0456. MAN and EEN acknowledge the ARO grant W911NF-14-1-0639. ANP acknowledges the support by the Government of the Russian Federation (Grant No. 074-U01) and the "Dynasty" Foundation. EEN was supported by ARO MURI, NSF Center for Photonics and Multiscale Nanomaterials, and Gordon and Betty Moore Foundation.

Figure 1.

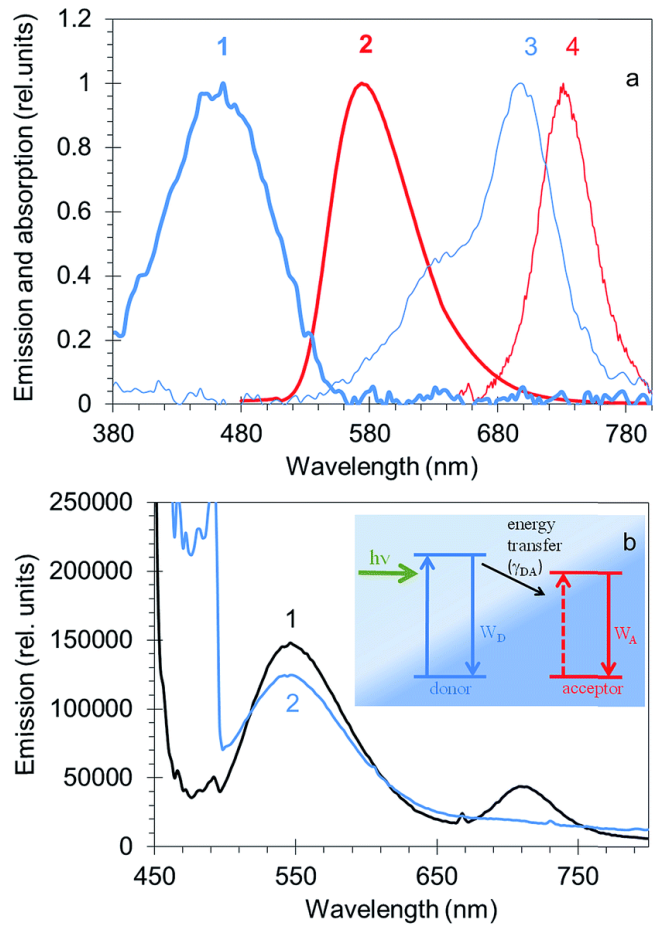

Figure 1: (a) Absorption (1,3) and emission (2,4) spectra of DCM (1,2) and DOTC (3,4) dyes doped into PMMA. (b) Emission spectra of ≈80 nm thick PMMA films co-doped by DCM and DOTC molecules (concentrations ≈ 33 mM DCM and 20 mM DOTC) deposited onto glass (1) and silver film (2) substrates and pumped at λ=400 nm into the absorption band of DCM. (Similar effect was observed on top of metamaterial with Ag as the outmost layer.) Inset: Schematic of Förster donor-acceptor energy transfer showing absorption of light in donor, combination of radiative and non-radiative relaxation processes in donor ($W_D$), donor-acceptor energy transfer ($\gamma_{DA}$), and the relaxation processes in acceptor ($W_A$).

Figure 2.

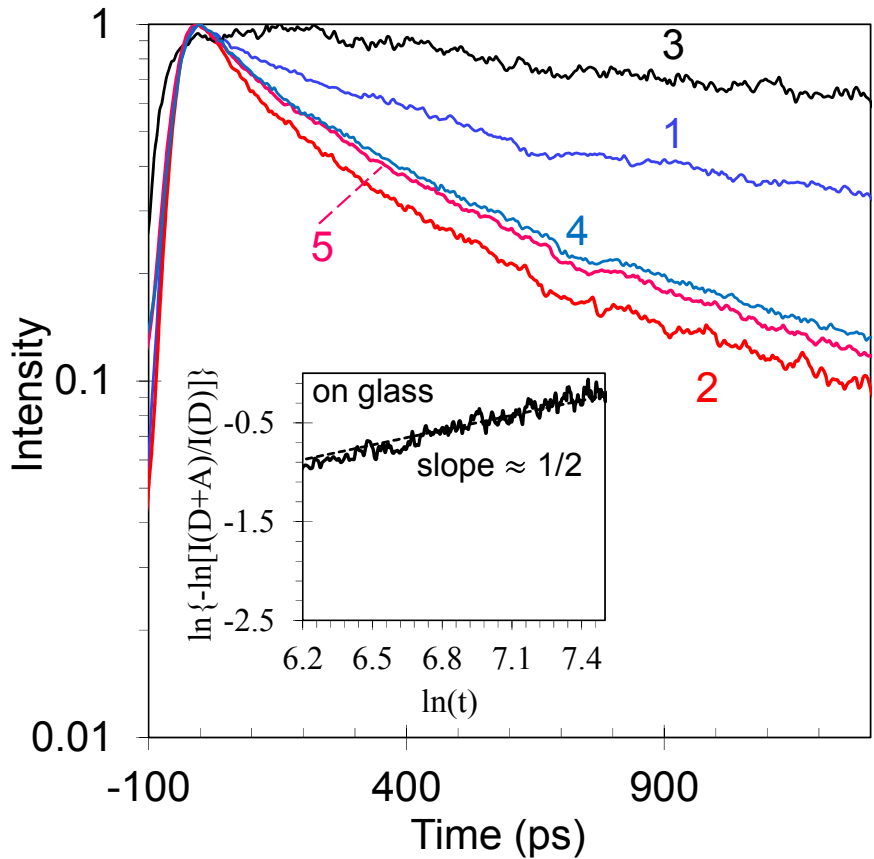

Figure 2: Spontaneous emission kinetics in PMMA films deposited on glass: (1) donors (DCM) in a singly doped PMMA film, (2) donors in the PMMA film co-doped with donors and acceptors, and (3) acceptors (DOTC) in the PMMA film co-doped with donors and acceptors. Spontaneous emission kinetics in PMMA films deposited on metamaterial substrates with Ag as the top layer: (4) donors (DCM) in a singly doped PMMA film and (5) donors in the PMMA film co-doped with donors and acceptors. All films were pumped into the absorption band of acceptors at $\lambda=392$ nm. Inset: the ratio of emission kinetics 2 and 1 from the main frame, showing slope ~1/2 when plotted as $\ln(-\ln(I(t)))$ vs $\ln(t)$. (Solid line indicates the slope equal to 1/2.)

Figure 3.

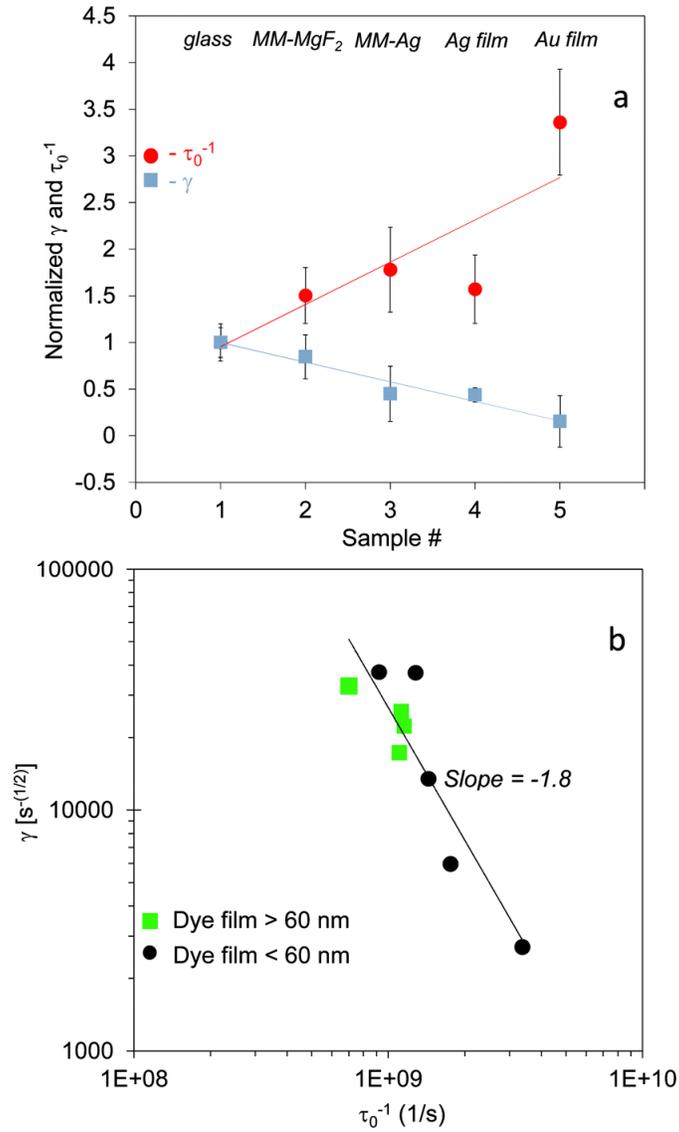

Figure 3: (a) Emission decay rates $\tau_0^{-1}$ and Förster energy transfer constants $\gamma$ in dye-doped films deposited on top of glass (1), metamaterial with $MgF_2$ as the outermost layer (2), metamaterial with Ag as the outermost layer (3), Ag film (4), and Au film (5). All data points are normalized to that on glass. (b) Values $\tau_0^{-1}$ plotted against corresponding values $\gamma$ in the dye-doped PMMA films deposited on the top of samples 1, 2, 3 and 4 in Fig. a.

Figure 4.

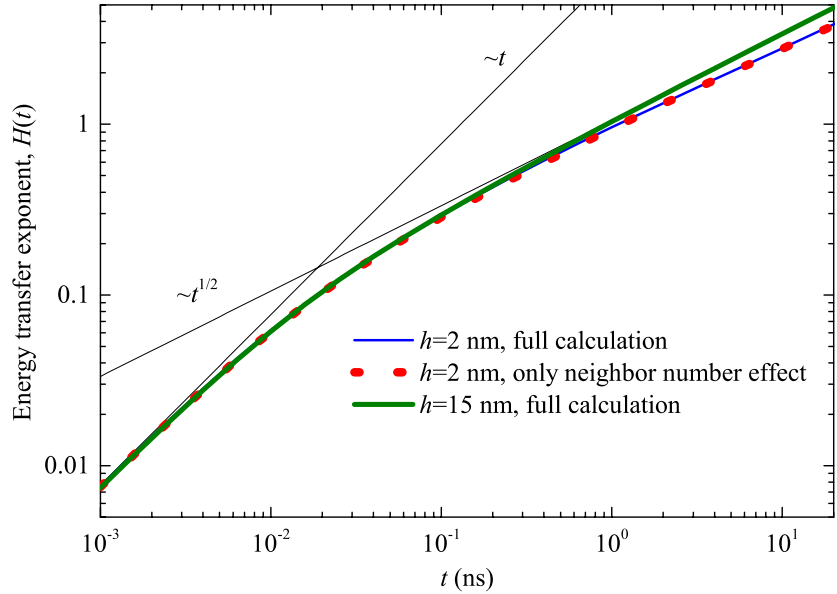

Figure 4: Green and blue solid curves have been calculated for the donor height over the Ag substrate $h$ equal to 15 nm and 2 nm, respectively. Red dotted curve has been calculated for $h=2$ nm neglecting the modification of the Green function due to the substrate. The calculation has been performed for the following set of parameters: $W_0 = 10^9$ s$^{-1}$, $\varepsilon_b = 2$, $n_A = 1.2 \times 10^{19}$ cm$^{-3}$, $d_{dye} = 30$ nm, $\hbar\omega = 2.1$ eV, $r_{min}=1$ nm (see Methods). Data from Ref. [37] was used for the dielectric permittivity of Ag. Thin black lines are guides for eye and have been obtained as linear fits with the slopes $H(t) \propto t$ and $H(t) \propto t^{1/2}$, respectively.

Figure 5.

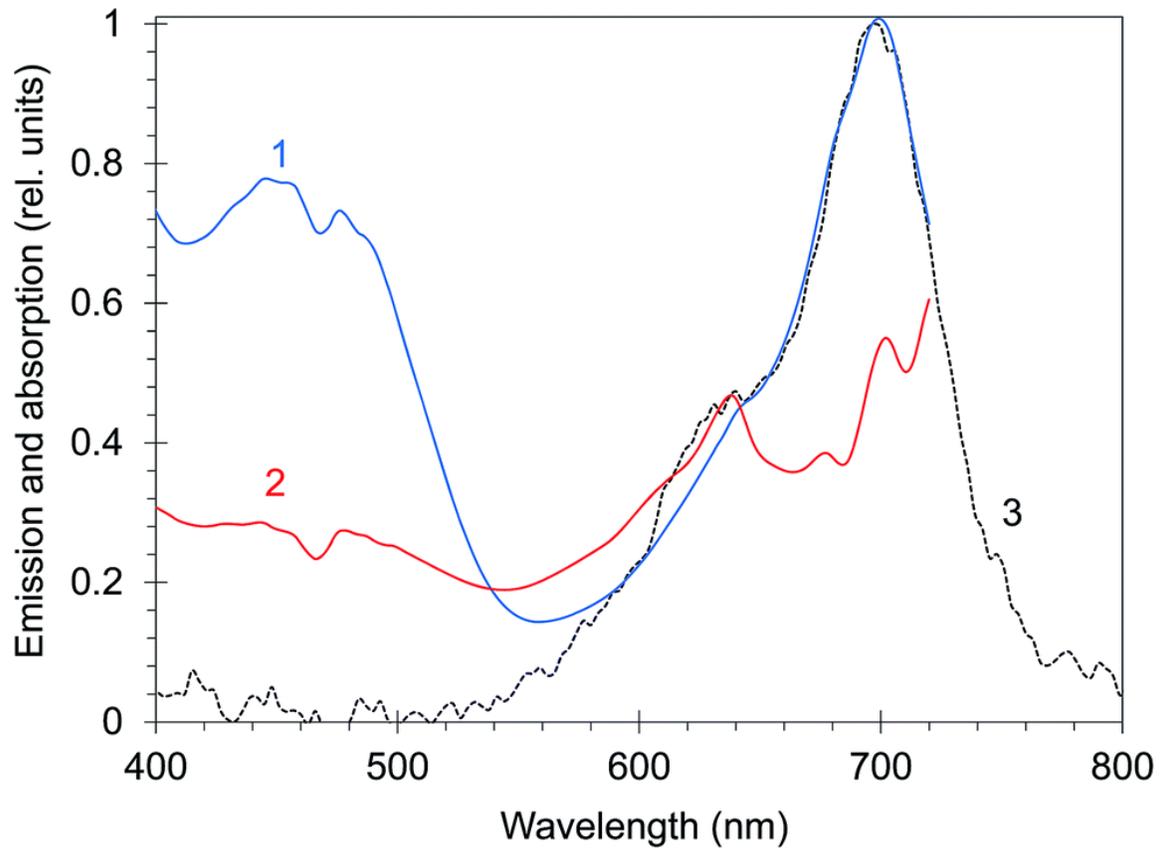

Figure 5: Excitation spectra of the DOTC molecules in the PMMA film (co-doped by the DCM dye) deposited on glass (trace 1) and Ag/MgF$_2$ metamaterial (trace 2). Dashed line – absorption spectrum of DOTC doped PMMA film deposited on glass.

Figure 6.

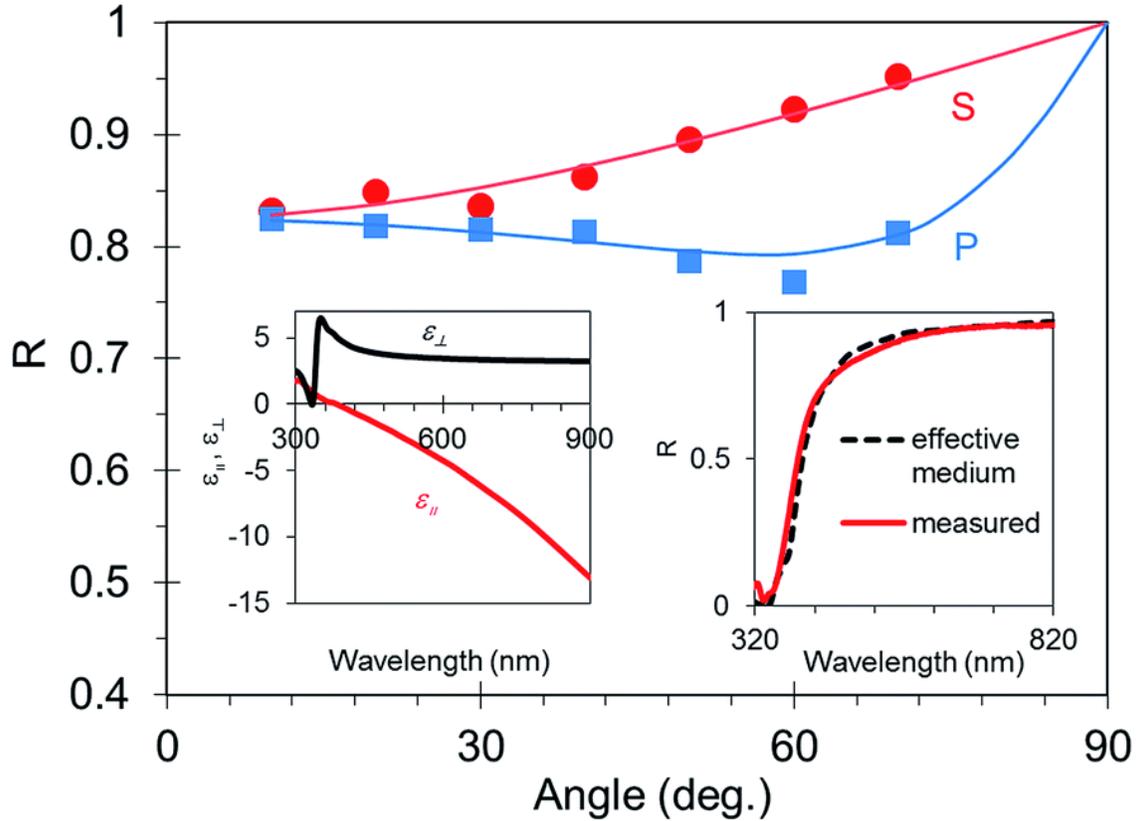

Figure 6: Angular reflectance spectra of a typical Ag/MgF$_2$ multilayered sample (Ag as the top layer) in s and p polarizations, at $\lambda$=543 nm. Markers – experiment, solid lines – fitting. Left inset: Effective medium spectra of real parts of dielectric permittivities $\varepsilon_\parallel$ and $\varepsilon_\perp$, calculated for a lamellar Ag/MgF$_2$ structure with Ag filling factor equal to 39%. Characters – experimentally measured values at $\lambda$=543 nm. Right inset: Reflectance spectrum of the lamellar Ag/MgF$_2$ sample. Solid line: experiment; dashed line: prediction of the effective medium theory, corresponding to a metal fill fraction of 39%.

Figure 7.

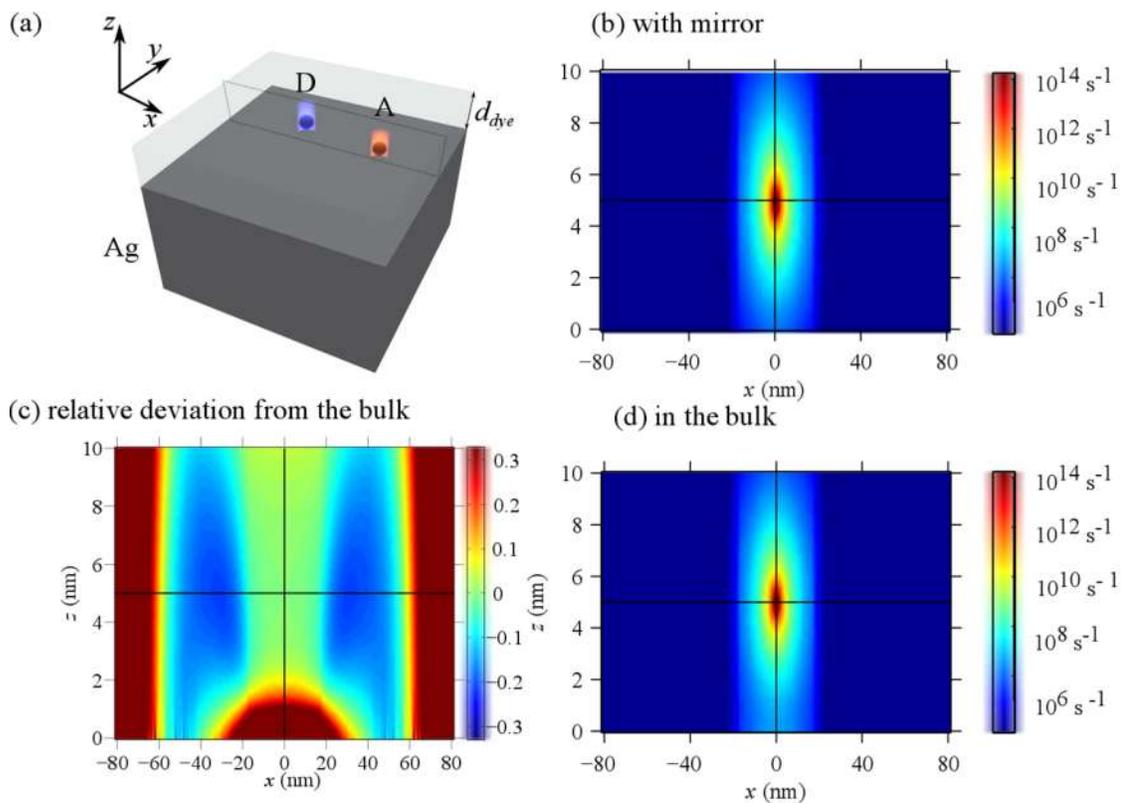

Figure 7: (a) Schematic illustration of the geometry of the problem. (b), (d) Transfer rate maps in the xz plane for the donor located above the mirror and in the bulk, respectively. (c) The energy transfer rate map, normalized to the bulk value as $W_{DA}/W_{DA}^{bulk} - 1$. The calculation has been performed for $r_D = h\hat{z}$ with $h = 5$ nm.

Figure 8.

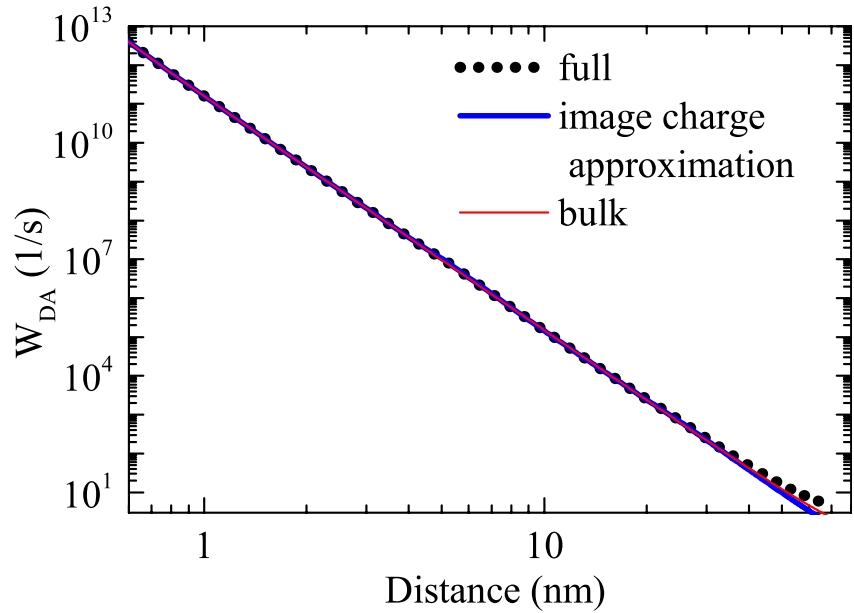

Figure 8: Dependence of the angular-averaged energy transfer rate versus donor-acceptor distance $r_{DA}$. Full calculation – dotted line, calculation in the electrostatic image-charge approximation – thick solit blue line, and calculation with the bulk Green function that does not include the effect of the mirror into account – thin solid red line. The donor has been positioned at $r_D = h\hat{z}$ with $h$ = 5 nm, other parameters are the same as in Fig. 4.

Figure 9.

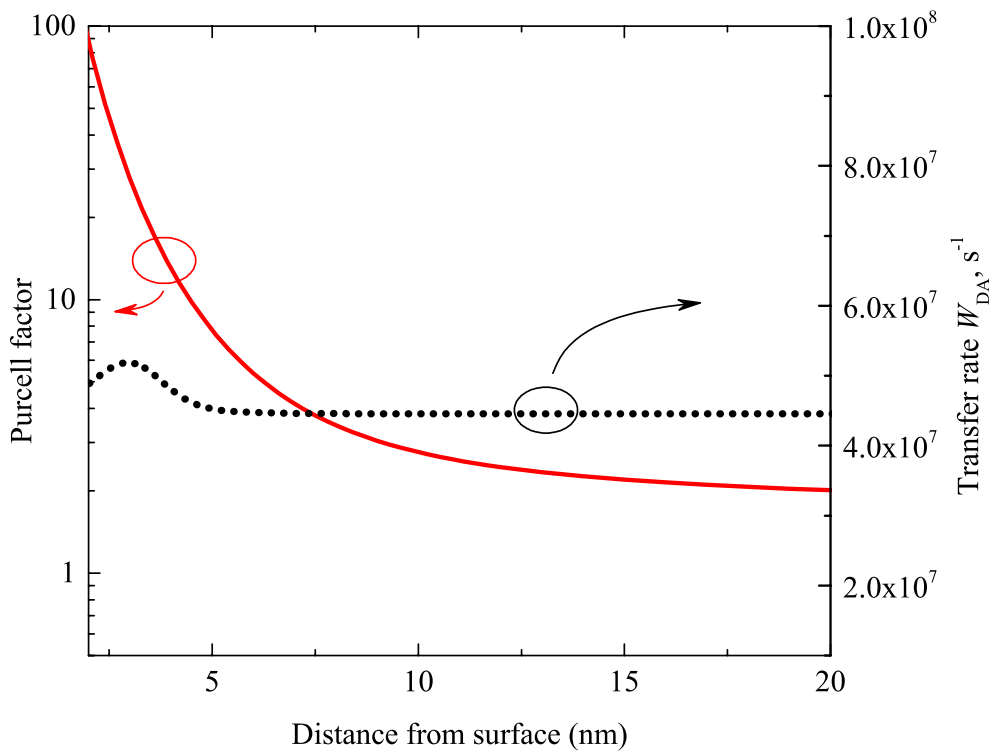

Figure 9: Comparison of the effec of the silver substrate on the Purcell factor and on the Energy transfer rate. Red solid curve shows the polarization-averaged Purcell factor as function of the height of the donor above the silver surface. Thick dotted curve shows the height dependence of the energy transfer rate. The transfer rates has been evaluated at donor-acceptor distances $r_{DA}$=4 nm, and averaged over angular distribution of acceptors. The calculation parameters are the same as in Fig. 4.